%% file: 0paper.tex
\documentclass[sigconf]{acmart}

\renewcommand\footnotetextcopyrightpermission[1]{} 
\setcopyright{none}

\usepackage{subfig}
\usepackage{algorithm}
\usepackage{algorithmic}
\usepackage{appendix}
\usepackage{multirow}
\usepackage{colortbl}
\usepackage[most]{tcolorbox}
\usepackage{pifont}
\usepackage{tikz}
\usepackage{enumitem}
\usepackage{wasysym}
\usepackage{xspace}
\usepackage{amsmath}
\newcommand{\method}{{\scshape Gecko}}
\newcommand{\cmark}{\CIRCLE}
\newcommand{\xmark}{\Circle}
\newcommand{\smark}{\LEFTcircle}

\newif\ifcomment
\commenttrue
\ifcomment
\newcommand{\virat}[1]{{\color{red} \textbf{Virat: #1}}}

\else
\newcommand{\virat}[1]{}
\fi

\begin{document}

\title{Towards Privacy Aware Deep Learning for Embedded Systems}

\author{Vasisht Duddu}
\affiliation{%
 \institution{Univ Lyon, INSA Lyon, Inria, CITI}
}
\authornote{Work partly done at author's current affiliation with University of Waterloo, Canada}
\email{vasisht.duddu@uwaterloo.ca}

\author{Antoine Boutet}
\affiliation{%
	\institution{Univ Lyon, INSA Lyon, Inria, CITI}
}
\email{antoine.boutet@insa-lyon.fr}

\author{Virat Shejwalkar}
\affiliation{%
	\institution{University of Massachusetts Amherst}
}
\email{vshejwalkar@cs.umass.edu}

\begin{abstract}
Memorization of training data by deep neural networks enables an adversary to mount successful membership inference attacks. 
Here, an adversary with \textit{blackbox} query access to the model can infer whether an individual's data record was part of the model's sensitive training data using only the output predictions. This violates the data confidentiality, by inferring samples from proprietary training data, and privacy of the individual whose sensitive record was used to train the model.
This privacy threat is profound in commercial \textit{embedded systems} with on-device processing. 
Addressing this problem requires neural networks to be \textit{inherently private by design} while conforming to the memory, power and computation constraints of embedded systems. This is lacking in literature. 

We present the first work towards membership privacy by design in neural networks while reconciling \textit{privacy-accuracy-efficiency} trade-offs for embedded systems.
We conduct an extensive \textit{privacy-centred} neural network design space exploration to understand the membership privacy risks of well adopted state-of-the-art techniques: model compression via pruning, quantization, and off-the-shelf efficient architectures. 
We study the impact of model capacity on memorization of training data and show that compressed models (after retraining) leak more membership information compared to baseline uncompressed models while off-the-shelf architectures do not satisfy all efficiency requirements. Based on these observations, we identify quantization as a potential design choice to address the three dimensional trade-off.

We propose \method~ training methodology where we explicitly add resistance to membership inference attacks as a design objective along with memory, computation, and power constraints of the embedded devices. 
We show that models trained using \method~ are comparable to prior defences against blackbox membership attacks in terms of accuracy and privacy while additionally providing efficiency. This enables \method~ models to be deployed on embedded systems while providing membership privacy. 
\end{abstract}


\maketitle
\pagestyle{plain}

\input{1introduction}

\input{2background}
\input{3setup}

\input{4analysis}
\input{5design}
\input{6evaluation}
\input{7related}

\input{8conclusions}

\bibliographystyle{ACM-Reference-Format}
\bibliography{paper}

\end{document}

%% file: 1introduction.tex
\section{Introduction}\label{introduction}

The tremendous performance of machine learning (ML), especially deep neural networks (NNs), has resulted in their deployment to low-powered wearable devices and embedded systems for various applications such as video processing and analytics~\cite{10.1145/3394171.3418551}.
Specifically, Internet of Things (IoT) devices prefer on-device processing to reduce communication latency and overhead, while also preserving the confidentiality of data from an untrusted data curator~\cite{8110880}.
Such NN architecture designs should conform to efficiency constraints on memory, energy, and computation for embedded devices, and also maintain high prediction accuracy. Additionally, privacy laws, such as HIPAA and GDPR, require on-device processing to maintain the privacy of an individual's sensitive data (e.g., medical records, location traces, purchase preferences) against privacy attacks such as membership inference attacks~\cite{shokri2017membership,salem2018ml,song2020systematic,8429311}. Hence, NNs deployed for low powered embedded systems are required to incorporate privacy, accuracy and efficiency.

\noindent\textbf{\underline{Membership Inference Attack.}} 
NNs, deployed for wearable devices and embedded systems (e.g., for remote patient monitoring), are trained using private and potentially sensitive data collected from multiple individuals. 
Such systems, executing the fully trained model, are then used by customers, for instance, to monitor health indicators from physiological signals.
An adversary, given access to the model via an API, leverages membership inference attack to identify whether a particular target data record was used in the proprietary and sensitive training data of the model. 
This reveals the health status of the individual corresponding to the target data record. 
This is a privacy risk in cases where the knowledge of a data record's membership in the model's training dataset is sensitive which has been extensively studied in prior literature~\cite{shokri2017membership,salem2018ml,song2020systematic,8429311}.
For instance, an adversary can exploit a model trained on data records containing diagnostic information of cancer patients to identify whether an individual has cancer based on if the data record was used to train the model.
This violates (a) the individual's privacy and (b) confidentiality of the sensitive data. In such cases, it is crucial to design NNs resistant to membership inference attacks, where the adversary infers unobservable, sensitive information (e.g., individual's health status) from the observable information (e.g., model predictions).

\noindent\textbf{\underline{Motivation: Privacy by Design.}} Embedded systems demand on-device processing of data using NNs while conforming to the memory, power and computation constraints, leading to an efficiency and accuracy trade-off~\cite{rastegari2016xnornet}. 
To bring NNs to edge devices, several optimizations such as model compression through pruning, quantization, and efficient off-the-shelf architectures are adopted. Currently, the designs of NNs for embedded systems do not consider privacy and such optimization schemes are not resistant to membership inference attacks, leaking private information of individuals' training data. 
Further, it has been shown that membership inference attacks are due to memorization of training data by large capacity NN models and current privacy defences result in \textit{privacy-accuracy} trade-off~\cite{Abadi:2016:DLD:2976749.2978318,DBLP:conf/ccs/NasrSH18}. However, \textit{the impact of model capacity on membership privacy is unexplored and understanding this is crucial to design NNs while accounting for privacy-efficiency-accuracy trade-off}. This problem of designing NNs for embedded systems which are inherently private by design along with good efficiency and accuracy, motivates our work. We address this by answering two research questions in sequence:
\begin{enumerate}[label=\textbf{RQ\arabic*},leftmargin=*]
    \item \label{req1} How does varying model capacity, which influences the memorization of training data, impact membership privacy risks?
    \item \label{req2} How can we design NNs with membership privacy along with accuracy and efficiency constraints for embedded systems?
\end{enumerate}

\noindent\textbf{\underline{Overview of Proposed Approach.}} First, we address \ref{req1} by studying the membership privacy risks of three state-of-the-art hardware software co-design techniques, i.e., namely, model compression, quantization and efficient off-the-shelf architectures. 
We show that model compression (i.e., pruning the network followed by retraining) leaks more information compared to baseline uncompressed models indicating a higher membership privacy risk to the individual’s data. Off-the-shelf architectures (i.e., MobileNet and SqueezeNet) do not meet all the efficiency requirements but limits the membership privacy risk. Quantization satisfies privacy and efficiency constraints at the cost of accuracy. 

This analysis forms the basis of identifying the best design choice for neural network architecture for embedded devices to address \ref{req2}.
We identify \textit{quantization} as our design choice. We propose \method~ — a two phase training methodology for designing NNs optimized specifically for privacy, accuracy and efficiency. Phase I of \method~ quantizes the model parameters and intermediate layerwise outputs by constraining their values to \{-1,+1\} to reduce the memory, energy consumption, and computation overhead. This optimizes the model for efficiency \textit{and} privacy but at the cost of accuracy.

\noindent\textbf{\underline{Main Contributions.}} We claim the following main contributions:
\begin{itemize}[leftmargin=*]
    \item We present the \textbf{first work on membership privacy by design in NNs} and conduct a privacy-centred design space exploration of NNs to understand the trade-offs between privacy, efficiency and accuracy for embedded devices. We find that quantization provides significant resistance to membership inference attacks while satisfying efficiency requirements compared to model compression and off-the-shelf efficient architectures. (Section~\ref{eval-efficiency} and~\ref{eval-leakage})
    \item We propose \method~ training for NNs by choosing \textbf{quantization as the key design choice combined with knowledge distillation} to improve the privacy, accuracy and efficiency trade-off\footnote{Code: \url{https://github.com/vasishtduddu/EmbeddedMIA}}. (Section~\ref{design} and~\ref{sec:eval})
    \item We show \method~ models are \textbf{comparable in accuracy and privacy risks to prior state-of-the-art defences against blackbox membership inference attacks}, adversarial regularization and differential privacy, while additionally providing efficiency for deployment to embedded systems. (Section~\ref{eval-defences})
\end{itemize}

%% file: 2background.tex
\section{Background}\label{background}

ML algorithms learn a function $f:X \rightarrow Y$ mapping from the input space $X$ to the space of corresponding class labels $Y$.
This is modeled as an optimization where the objective is to find the parameters $\theta$ by minimizing the model's loss, $min_{\theta} L(f(x),y;\theta)$.
We discuss the constraints of executing NNs on embedded devices (Section~\ref{edl}) and the current state-of-the-art NN designs tailored for embedded devices (Section~\ref{designs}). Then, we describe the privacy threat of membership inference attacks considered in this work (Section~\ref{threatmodel}).

\subsection{Requirements for Embedded Devices}\label{edl}

The list below presents the most important requirements accounted by designers of NNs based embedded systems~\cite{8114708}, and unfortunately, privacy is not a part of it:

\begin{itemize}[leftmargin=*]
\item \textbf{Energy Efficiency.} Energy consumption is a vital constraint to maximize the battery lifetime of low powered embedded devices which operate for long duration.
In NN execution, multiply and accumulate (MAC) operations computes the product between model parameter matrix and layerwise output vector (called activations) and stores the intermediate results. Specifically, MAC operations access memory to read weights, inputs, and intermediate outputs from previous layer, and write the computed output to the memory. These read-write operations consume energy which is higher than the actual MAC computations~\cite{6757323}.
Hence, energy efficiency is achieved by reducing the memory accesses by (a) optimizing hardware to exploit sparsity in MACs~\cite{Umuroglu2017FINNAF} and (b) reducing the precision to increase the throughput of data~\cite{rastegari2016xnornet,Li2016TernaryWN,NIPS2016_6573}.

\item \textbf{Computation Efficiency.} The total MAC operations quantifies the requirement of computation efficiency~\cite{8114708}.
The computation speed of CPUs on embedded devices restricts the processing rate of MACs, which reduces with the number of parameters (of NNs)~\cite{rastegari2016xnornet,Li2016TernaryWN,NIPS2016_6573}.
Additionally, replacing MACs with cheaper binary arithmetic significantly lowers the computational overhead~\cite{rastegari2016xnornet,Li2016TernaryWN,NIPS2016_6573}.

\item \textbf{Memory Efficiency.} The memory footprint of NN parameters and additional runtime storage for intermediate outputs, should be within the memory constraints of the embedded devices.
There are two ways to achieve this requirement:
(a) reducing the precision of the parameters and intermediate outputs~\cite{rastegari2016xnornet,Li2016TernaryWN,NIPS2016_6573} and (b) pruning the parameters by increasing sparsity~\cite{DBLP:journals/corr/HanMD15,journals/corr/YangCS16a,DBLP:journals/corr/HanPNMTECTD16,Han:2015:LBW:2969239.2969366}.
\end{itemize}

\subsection{Neural Network Designs for Efficiency}\label{designs}

To address the computational, memory and energy requirements of embedded devices, model designers use three state-of-the-art approaches for designing efficient NN models for embedded systems: (a) model compression via pruning, (b) quantization of model parameters and activations and (c) designing standard architectures (off-the-shelf efficient architectures).

\noindent\textbf{\underline{Model Compression (Pruning + Retraining).}}\footnote{We use the terms pruning and compression interchangeably}
NNs are overparameterized, i.e., have a large number of redundant weights.
Pruning the network removes these redundant weights (setting them as ``0'') without degradation of model accuracy.
The pruning operation results in a model with sparse parameters for which the hardware designers skip the multiplication and memory storage to improve efficiency.
Sparse weights can be stored in a compressed format in the hardware using the compressed sparse row or column format which reduces the overall memory bandwidth~\cite{DBLP:journals/corr/HanMD15,10.1109/ISCA.2016.30,Han:2015:LBW:2969239.2969366}.
Furthermore, aggressive pruning to compress the model significantly, requires the model to be re-trained to restore the original accuracy.
For a sensitivity threshold $\tau$ and NN model $f_{\theta}$, the parameters $\theta$ close to ``0'' are replaced by ``0'': 
$
    f'(\theta)=
\begin{cases}
    0, & \text{if } -\tau \leq \theta \leq \tau\\
    \theta,  & \text{otherwise}
\end{cases}
$

\noindent\textbf{\underline{Off-the-Shelf Efficient Architectures.}}
NNs can be redesigned by changing the hyperparameters (i.e., filter size in convolution, number of layers and their types) to reduce the number of parameters and hence, the memory footprint.
One approach is to replace larger convolution filters with multiple smaller filters with less number of parameters but covering the same receptive fields.
For instance, one 5x5 filter can be replaced by two 3x3 filters.
Alternatively, 1x1 convolutional layers reduce the number of channels in output feature map, lowering the computation and number of parameters.
For instance, for an input activation of dimension 1x1x64, 32 1x1 convolutional filter downsamples the activation maps to get an output of 32 channels.
Such optimizations enable to design compact network architecture with layers having lower parameters compared to the original model, extensively adopted in MobileNet~\cite{conf/cvpr/SandlerHZZC18} and SqueezeNet~\cite{DBLP:journals/corr/IandolaMAHDK16}.

\noindent\textbf{\underline{Quantization.}}
Quantization reduces the precision of the model's parameters and the intermediate activations during execution.
Quantization maps parameters and activations to a fixed set of discrete levels~\cite{Hubara:2017:QNN:3122009.3242044}.
The number of quantized levels determines the precision of the operands ($log_2(\#levels)$).
This reduction in the precision of the parameters lowers the storage cost of the model in memory. 
In addition, reducing the precision of activations lowers the computation overhead by replacing MACs with binary arithmetic. This further reduces the energy consumption by lowering the memory accesses and increasing throughput.
Aggressively quantizing the parameters and activations to binary and ternary precision significantly improves the overall efficiency, however, at the cost of accuracy~\cite{rastegari2016xnornet}.
For instance, Binarized NNs quantize the operands to \{-1,+1\} values~\cite{NIPS2016_6573} while ternary NNs have values \{-$w$, 0, $w$\} where $w$ can be fixed or learnt during training~\cite{Li2016TernaryWN}. These are examples of uniform quantization.

Alternatively, weight sharing maps several parameters to a single value reducing the number of unique parameters~\cite{DBLP:journals/corr/HanMD15}.
This mapping is done using K-Means clustering or a hashing function where a \emph{codebook} maps different parameters to the corresponding shared values.

\subsection{Privacy Threat: Membership Inference}\label{threatmodel}

NNs for embedded systems do not account for privacy threats, however these threats still exist if personal and potentially sensitive data feed the system.
For instance, if the adversary learns something specific about a user's data record used in the training dataset, we refer to such information as membership privacy leakage.
This privacy leakage about a user's record can be, for instance, the membership details of the record in the training set of the model, referred to as membership inference attack.

In this work, we specifically use membership inference attacks to quantify information leakage in ML models following several prior work~\cite{shokri2017membership,salem2018ml,song2020systematic,8429311}.
ML models are more confident while predicting the class of already seen train data records compared to unseen test data records.
Membership inference attacks exploit this difference in the model's confidence to classify a new data record as being a ``Member'' or ``Non-Member'' of the model's training data.
This is a binary decision problem where the adversary classifies the membership of a given input $x$ using the model's output prediction $f(x;\theta)$ to infer whether a given data record was used in the model's training data or not.

\noindent\textbf{\underline{Attack Details.}} Given a user's data record $x$ $\sim$ $P(X,Y)$, where $P(X,Y)$ is the data distribution from which the training data $D_{train}$ was sampled, the adversary estimates $P(x \in D_{train})$ using the model's prediction $f(x;\theta)$.
Empirically, the adversary identifies a threshold to estimate whether $x \in D_{train}$ which can also be learnt using a binary classifier.
In this work, we use the confidence score attack where the adversary obtains $f(x;\theta)$ and finds the maximum posterior value and infers $x \in D_{train}$ if the value is greater than a threshold~\cite{salem2018ml,8429311}.
The attack is based on the observation that the maximal posterior of a member data record is higher (more confident) than a non-member data record of the training dataset. This particular attack has been used in several prior work for empirically estimating membership privacy risks~\cite{song2020systematic,salem2018ml,duddu2020quantifying,8429311,10.1145/3319535.3354211}.

\noindent\textbf{\underline{Threat Model.}} We consider a blackbox setting where the adversary is assumed to have no knowledge about the target model.
Formally, given a target model $f()$, the adversary only sees the final model prediction $f(x;\theta)$.
The adversary does not know the architecture of $f()$ nor the model parameters $\theta$.
We do not consider whitebox setting where the adversary has the access to both the model output predictions $f(x;\theta)$ as well as the architecture of $f()$ and the model parameters $\theta$.
Indeed, this whitebox setting does not necessarily result in any benefit to the adversary in terms of attack accuracy (shown theoretically~\cite{pmlr-v97-sablayrolles19a} and empirically~\cite{shejwalkar2020membership,song2020systematic}).
Consequently, blackbox setting is a more practical setting seen typically in ML as a Service (MLaaS)and commercial embedded devices such as wearable and IoT devices, where the adversary submits an input query to the trained model via an API and obtains the corresponding output.

%% file: 3setup.tex
\section{Experiment Setup}\label{compare}

We describe our experimental setup: the datasets and architectures used in our analysis (Section~\ref{datasets}), and the considered metrics (Section~\ref{metrics}).

\subsection{Datasets and Architectures}\label{datasets}

For evaluating and comparing different efficiency algorithms, we mainly use four standard benchmarking datasets: FashionMNIST, CIFAR10, PURCHASE, and LOCATION.
We train the model for 75 epochs for FashionMNIST and 100-150 epochs for CIFAR10, PURCHASE, and LOCATION.

\noindent\textbf{\underline{FashionMNIST}} consists of 60,000 training examples and a test set of 10,000 examples.
Each data record is a 28$\times$28 grayscale image which is mapped to one of 10 classes consisting of fashion products such as coat, sneaker, shirt, shoes.
For this dataset, we use a modified LeNet architecture with two convolution layers followed by maxpool and dense layers: [Conv 32 (3,3), Conv 64 (3,3), Maxpool (2,2), Dense 128, Dense 10] (Architecture 1). Additionally, we use a fully connected model [512, 512, 512] (Architecture 2).

\noindent\textbf{\underline{CIFAR10}} is a major image classification benchmarking dataset where the data records are composed of 32$\times$32 RGB images where each record is mapped to one of 10 classes of common objects such as airplane, bird, cat, dog.
For this dataset, we use standard state-of-the-art architectures: Network in Network (NiN), AlexNet and VGGNet.

\noindent\textbf{\underline{PURCHASE}} is a dataset capturing the purchase preferences of online customers taken from the authors of~\cite{shokri2017membership}.
The data records have 600 binary features and each record is classified into one of 100 classes identifying each user's purchase.
For this dataset, we use a fully connected architecture with the nodes in each layers as [1024, 512, 256, 128, 100].

\noindent\textbf{\underline{LOCATION}} is a dataset capturing user's location check-ins taken from the authors of~\cite{shokri2017membership} where each record has 446 binary features which is mapped to one of 30 classes each representing a location. For this dataset we use a fully connected architecture with hyperparameters as [512, 256, 128, 30].

\subsection{Metrics}\label{metrics}

We consider two types of metrics in our evaluation capturing the efficiency and the privacy.

\noindent\textbf{\underline{Efficiency.}} We evaluate efficiency in terms of memory, computation and energy.
Memory efficiency is compared based on the reduction in the memory footprint of the model computed from the parameters stored in the memory. Computation efficiency is compared based on the reduction in the MAC operations which influences the execution time. Finally, the energy consumption is compared based on memory accesses from reading inputs and writing results to the memory. Since, significant literature has compared the efficiency empirically, we provide a qualitative comparison for the baseline algorithms from prior work~\cite{8114708}. We instead focus on privacy-centred analysis as our main contribution building on top of the efficiency analysis from prior work.

\noindent\textbf{\underline{Privacy.}} We use the membership inference attack accuracy to estimate the membership privacy risks.
An accuracy above random guess $50\%$ indicates a training data membership leakage.
This indicates that the adversary is able to identify the membership status of a data record with an accuracy higher than random guess.
The success of inference attack accuracy is strongly correlated with the model's extent of overfitting empirically measured as the difference between the train and test accuracy (i.e., generalization error). Higher generalization error (i.e., overfitting) results in higher distinguishability between the test and train resulting in higher membership inference accuracy~\cite{shokri2017membership}.
Additionally, the accuracy of the model is computed on an unseen test data.

%% file: 4analysis.tex
\section{Efficiency Analysis}\label{eval-efficiency}

In view of the memory, computation and energy efficiency requirements, we qualitatively compare the three baseline algorithms (i.e., model compression, off-the-shelf architecture, and quantization).

\noindent\textbf{\underline{Memory Efficiency.}} We specifically focus on the impact of the three optimization techniques on the model's static memory required to store the model parameters. This static memory requirement is the major bottleneck compared to the runtime memory requirement (storing the intermediate activation outputs with lower dimension than model parameters). 

\begin{itemize}[leftmargin=*]
    \item \textbf{Model Compression}: Here, the model parameters which are pruned are simply replaced by a value of ``0''. Hence, storing even the ``0'' parameter takes up memory and does not necessarily decrease the overall memory footprint unless the hardware is optimized to skip the storage of all the zero values in the memory. This requires additional hardware logic to exploit the sparsity which may not be readily available.
    \item \textbf{Off-the-shelf Architectures}: The models are designed to specifically reduce the memory footprint. For instance, the memory footprint of Squeezenet and MobileNet is 5MB and 14MB compared to 250MB of Alexnet and more than 500MB of VGG architectures~\cite{DBLP:journals/corr/IandolaMAHDK16,conf/cvpr/SandlerHZZC18}.
    \item \textbf{Quantization}: Lowering the model precision from 64 or 32 bit floating point to binary precision results in a direct reduction of 64x or 32x in the overall memory footprint of the model.
\end{itemize}

\noindent\textbf{\underline{Computation Efficiency.}} A high computational efficiency, i.e., lower number of floating point MAC operations between parameters and intermediate outputs from each layer, typically results in faster inference.
\begin{itemize}[leftmargin=*]
    \item \textbf{Model Compression}: Here, achieving efficiency requires additional hardware optimization. Particularly, instead of actually computing the multiplications with ``0'' pruned values, the hardware optimization enables the user to skip the computation and replace the output by a ``0'' directly.
    \item \textbf{Off-the-shelf Architectures}: These models replace the complex matrix-vector multiplications to smaller dimensions. This reduces the overall number of parameters but it has been shown empirically\footnote{https://github.com/albanie/convnet-burden} that this does not necessarily reduce the number of multiply accumulate operations~\cite{article}.
    \item \textbf{Quantization}: For models with binarized parameters and activations, the MACs can be replaced by binary XNOR operations, maxpool can be replaced by OR operation, while the activations can be replaced by checking the sign bit. This reduces the floating point operations per second drastically~\cite{235489}.
\end{itemize}

\noindent\textbf{\underline{Energy Efficiency.}} Energy efficiency does not vary much with reduction of number of parameters and data type, but the number of memory accesses plays a vital role~\cite{6757323}. The benchmarking of energy consumption for different optimization and architectures is well explored and out of scope of this work. We refer the readers to~\cite{8114708} for more details. We qualitatively summarize the impact on energy consumption for the three specific algorithms considered.
\begin{itemize}[leftmargin=*]
    \item \textbf{Model Compression}: energy efficiency can be marginally improved by additionally providing hardware optimization~\cite{journals/corr/YangCS16a,DBLP:journals/corr/HanMD15}.
    \item \textbf{Off-the-shelf Architectures}: while computation efficiency improves, the energy efficiency is close to large scale state-of-the-art models like AlexNet~\cite{DBLP:journals/corr/IandolaMAHDK16,8114708}.
    \item \textbf{Quantization}: the energy efficiency is high as the memory access can be drastically reduced by increasing the throughput of data fetched from the memory. Specifically, lowering the precision from 32 bit floating point to binary results in lowering the memory accesses and 32x improvement in energy consumption~\cite{NIPS2016_6573,rastegari2016xnornet}. While some improvements are seen natively for quantized models (from replacing MACs with XNOR), higher benefits can be achieved via additional hardware optimization~\cite{Umuroglu2017FINNAF}.
\end{itemize}

\begin{tcolorbox}[text width=0.97\columnwidth,top=1pt,bottom=1pt,left=0pt,right=0pt,
 colback=gray!10!white,colframe=gray!10!white,colbacktitle=gray!80!black]
\underline{\textbf{Summary:}} A qualitative comparison of efficiency based on empirical results from prior work between different optimization techniques indicates that the quantized architectures show significant benefits over the other alternatives.
\end{tcolorbox}

\begin{figure*}[!htbp]
    \centering
    \subfloat[FashionMNIST]{{\includegraphics[width=4.5cm]{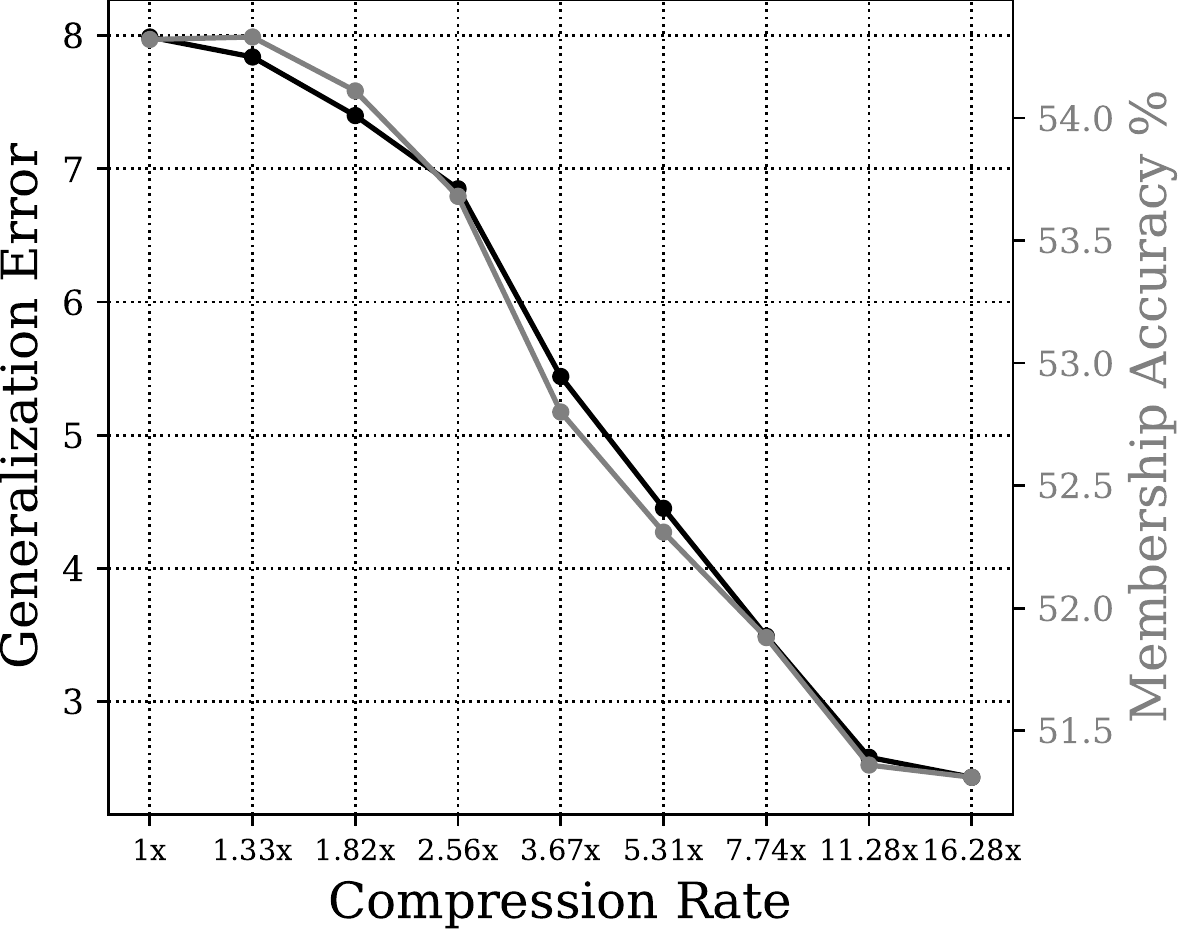} }}%
    \qquad
    \subfloat[LOCATION]{{\includegraphics[width=4.5cm]{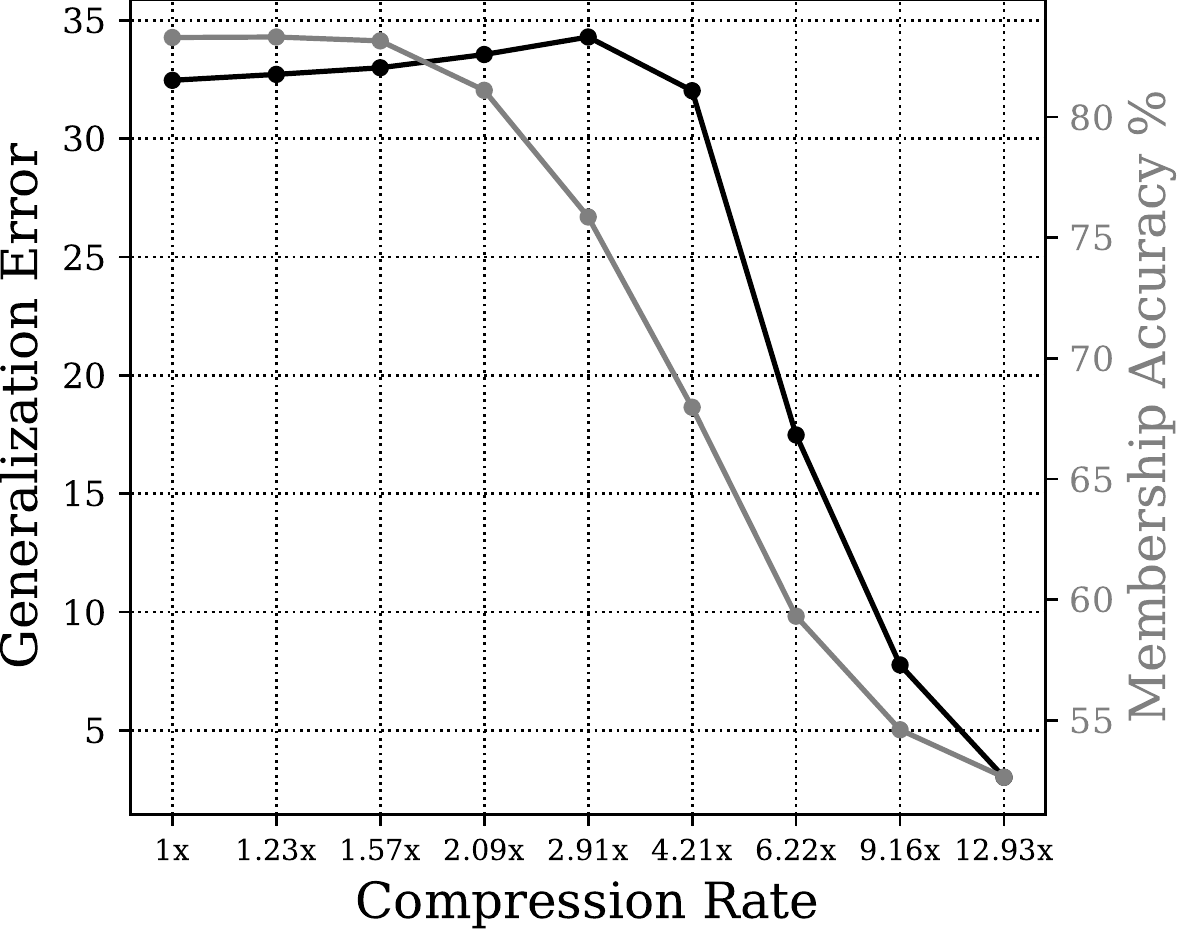} }}%
    \qquad
    \subfloat[PURCHASE]{{\includegraphics[width=4.5cm]{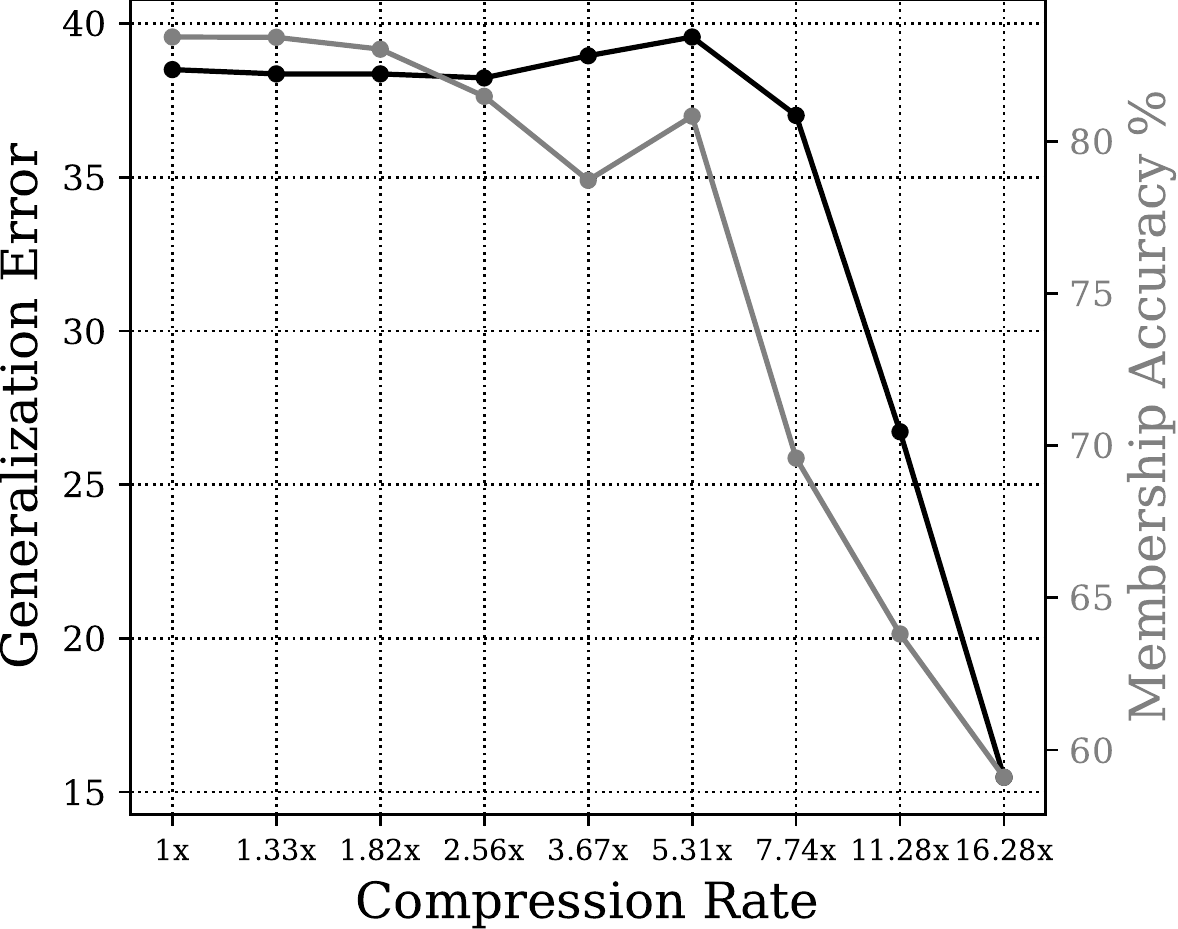} }}%
\caption{Impact of Pruning on Membership Privacy Risk: Membership privacy risk decreases due to decrease in generalization error for higher compression rates but at the cost of test accuracy.}
\vspace{-4mm}
\label{fig:xprune}
\end{figure*}

\begin{figure*}[!htbp]
    \centering
    \subfloat[FashionMNIST]{{\includegraphics[width=4.5cm]{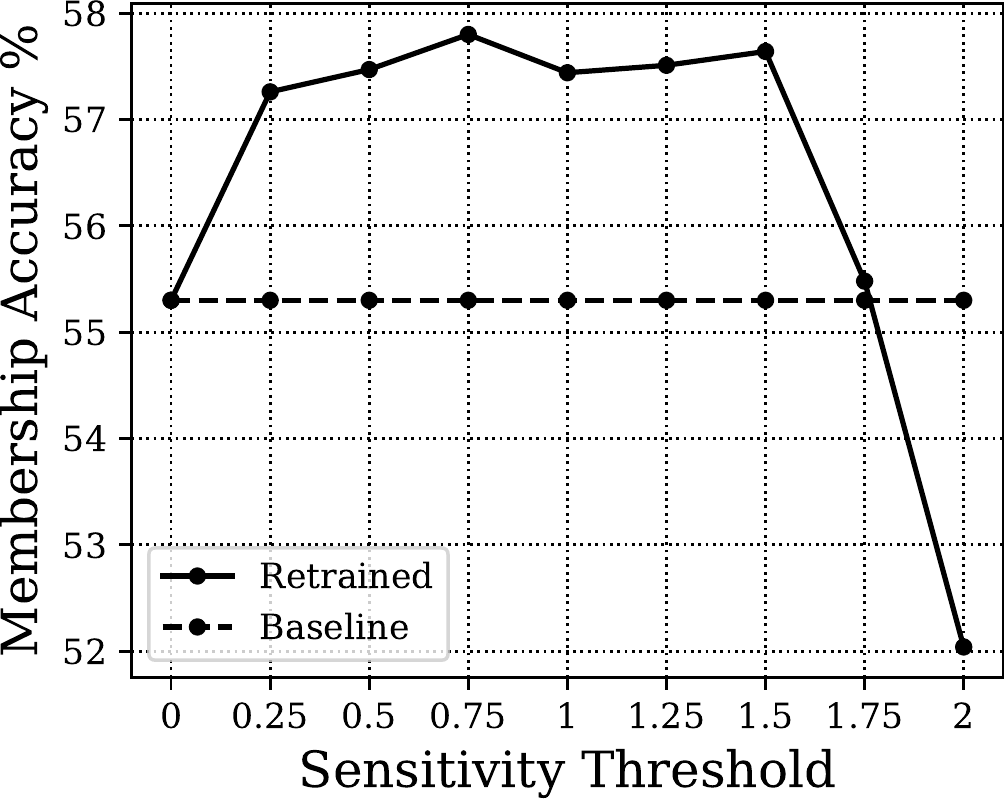} }}%
    \qquad
    \subfloat[LOCATION]{{\includegraphics[width=4.5cm]{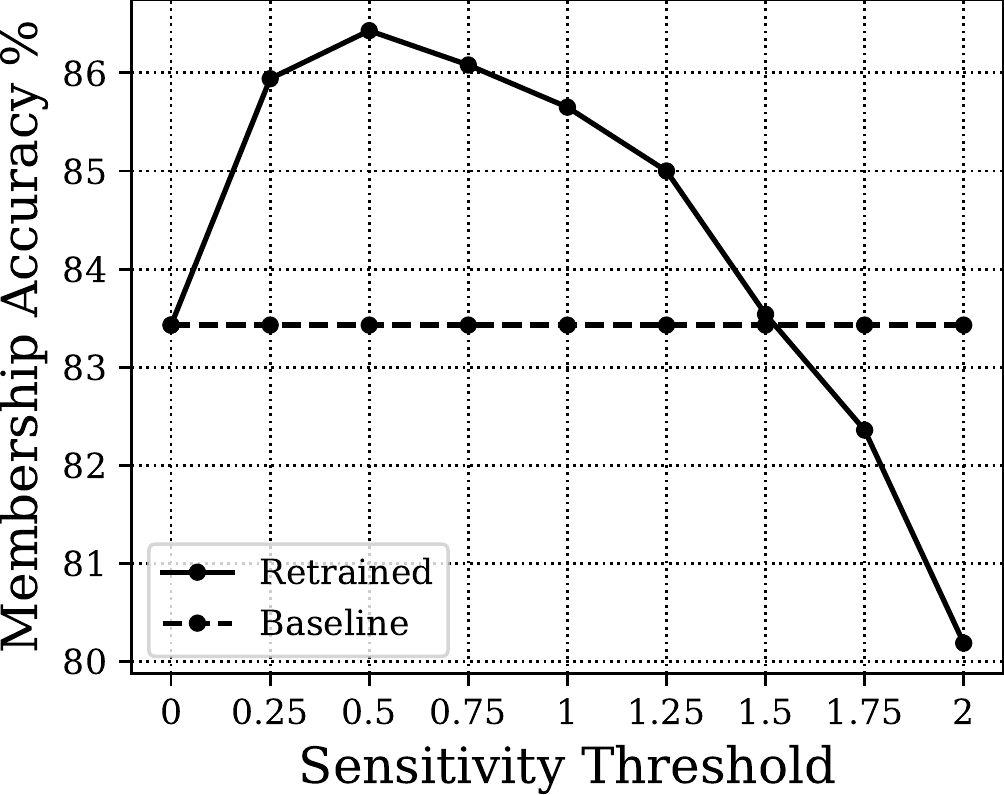} }}%
    \qquad
    \subfloat[PURCHASE]{{\includegraphics[width=4.5cm]{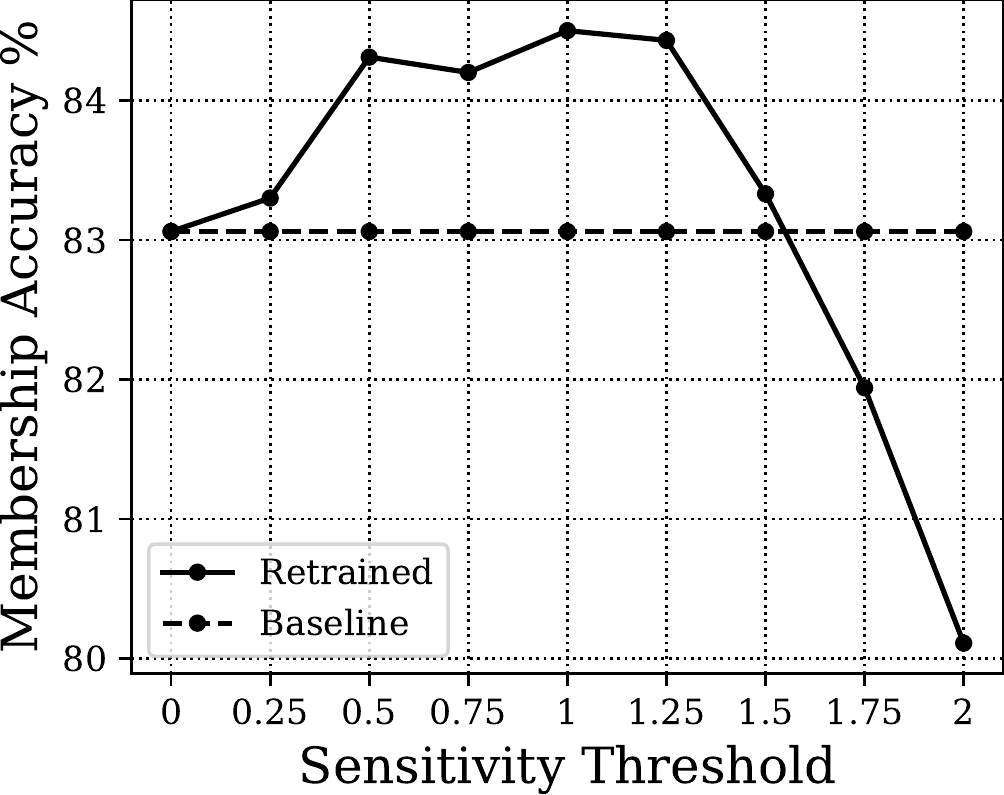} }}%
\caption{Impact of Retraining on Membership Privacy Risk: Retraining the pruned model shows a higher privacy risk than unpruned (baseline) model due to higher memorization of information per parameter.}
\label{fig:xretrain}
\vspace{-2mm}
\end{figure*}

\section{Privacy Analysis}\label{eval-leakage}

In this section, we address research question~\ref{req1} by evaluating the privacy leakage from membership inference attacks for model compression (Section~\ref{cmc}), off-the-shelf architectures (Section~\ref{cosa}), and quantization (Section~\ref{quant}).

\subsection{Model Compression}\label{cmc}

We evaluate the privacy leakage on compressing a model by pruning the connections (i.e., parameters corresponding to synapses between the nodes of two adjacent layers).
Here, pruning is achieved by replacing some of the parameters with ``0'' value.
As described in the original paper~\cite{Han:2015:LBW:2969239.2969366,DBLP:journals/corr/HanPNMTECTD16}, pruning is followed by retraining the model to restore the model's original accuracy with the reduced connections.
We evaluate and validate the impact on membership privacy on compressing the model trained on three datasets: FashionMNIST, LOCATION and PURCHASE.

\noindent\textbf{Impact of Pruning Parameters.} As the compression rate increases, the generalization error decreases (owing to a decrease in both train and test accuracy) which results in a decrease in membership attack accuracy (Figure~\ref{fig:xprune}).
This is expected as the model parameters are responsible for memorizing the training data information~\cite{DBLP:journals/corr/abs-1812-00910,236216,10.1145/3133956.3134077} and pruning the parameters lowers the memorization capacity of the model as well as the adversary's attack success. Hence, lowering the model capacity via compression indeed lowers the privacy risk to membership inference attacks at the cost of test accuracy.

\noindent\textbf{Impact of Retraining Pruned Model.} Interestingly, on retraining the pruned model for the initial sensitivity threshold values, we observe that the membership inference attack accuracy is much higher than the original unpruned baseline model (Figure~\ref{fig:xretrain}).
This indicates that the overall model compression algorithm (pruning + retraining) in turn increases the overall membership privacy leakage.
This can be attributed to the lower number of parameters which are forced to learn the same amount of information stored previously in the unpruned model with larger number of parameters.
In other words, the same amount of information is now captured by less number of parameters resulting in higher information stored per parameter~\cite{collins2017capacity}.
However, on aggressive pruning and retraining, the train and test accuracy also decreases resulting in a decrease in the information per parameter. This is empirically indicated by a decrease in membership inference attack accuracy after 1.5-1.75 sensitivity threshold of pruning (see Figure~\ref{fig:xretrain}).

\begin{tcolorbox}[text width=0.97\columnwidth,top=1pt,bottom=1pt,left=0pt,right=0pt,
 colback=gray!10!white,colframe=gray!10!white,colbacktitle=gray!80!black]
\underline{\textbf{Summary:}} Reducing model capacity by aggressively pruning parameters lowers the membership privacy risk but at the cost of test accuracy. Retraining the model to restore accuracy results in a higher membership privacy leakage compared to the baseline uncompressed model, making it a poor candidate for privacy sensitive applications. 
\end{tcolorbox}

\subsection{Off-the-Shelf Efficient Architectures}\label{cosa}

We now evaluate two popular state-of-the-art architectures, SqueezeNet and MobileNet, designed for low power systems trained on CIFAR10 dataset.
This evaluation is done only on CIFAR10 dataset (i.e., gathering a collection of images) as these state-of-the-art architectures are specialized for bigger datasets and not adapted for the smaller datasets.
As seen in Table~\ref{stdarch}, the SqueezeNet and MobileNet models show lower membership inference attack accuracy of 53.07\% and 55.57\% compared to larger models which have higher membership privacy risks.

\begin{table}[h]
\caption{Model capacity influences membership privacy risks.}
\vspace{-3mm}
\begin{center}
\renewcommand\arraystretch{1.5}
\fontsize{7.5pt}{7.5pt}\selectfont
\begin{tabular}{|c|c|c|c|c|}
\hline
\multicolumn{5}{|c|}{\textbf{CIFAR10}}\\
\hline
\textbf{Architecture} & \textbf{Memory} & \textbf{Train}  & \textbf{Test}  & \textbf{Inference}   \\
 & \textbf{Footprint} & \textbf{Accuracy} & \textbf{Accuracy} & \textbf{Accuracy}  \\
\hline
\textbf{SqueezeNet} & 5 MB & 88.21\% & 81.92\% & \cellcolor{green!25}53.07\% \\
\textbf{MobileNetV2} & 14 MB & 97.50\% & 87.24\% & \cellcolor{green!25}55.57\% \\
\hline
\textbf{AlexNet} & 240 MB & 97.86\% & 80.34\% & \cellcolor{red!25}60.40\% \\
\textbf{VGG11} & 507 MB & 99.13\% & 86.43\% & \cellcolor{red!25}58.04\% \\
\textbf{VGG16} & 528 MB & 99.58\% & 88.95\% & \cellcolor{red!25}58.70\%  \\
\textbf{VGG19} & 549 MB & 99.09\% & 88.18\% & \cellcolor{red!25}57.85\% \\
\hline
\end{tabular}
\end{center}
\label{stdarch}
\vspace{-3mm}
\end{table}



\begin{tcolorbox}[text width=0.97\columnwidth,top=1pt,bottom=1pt,left=0pt,right=0pt,
 colback=gray!10!white,colframe=gray!10!white,colbacktitle=gray!80!black]
\underline{\textbf{Summary:}} Lowering the model capacity by reducing the memory footprint indeed lowers privacy risk as privacy risks for off-the-shelf architectures compared to large capacity models. 
\end{tcolorbox}

\subsection{Quantization}\label{quant}

We now evaluate the technique of reducing the precision of both model's parameters and intermediate activations.
We consider the extreme and the most optimized case of binarizing the parameters and activations.
As a motivating example, we evaluate on FashionMNIST dataset for two model architectures with convolutional and fully connected layers as seen in Table~\ref{fmnist_quantize}.

\begin{table}[!htb]
\caption{Reducing the model precision decreases the inference attack but at the cost of test accuracy.}
\begin{center}
\renewcommand\arraystretch{1.5}
\fontsize{7.5pt}{7.5pt}\selectfont
\begin{tabular}{|c|c|c|c|c|}
\hline
\multicolumn{5}{|c|}{\textbf{FashionMNIST}}\\
\hline
\textbf{Architecture} & \textbf{Memory} & \textbf{Train}  & \textbf{Test}  & \textbf{Inference}  \\
 & \textbf{Accuracy} &  \textbf{Footprint} & \textbf{Accuracy} & \textbf{Accuracy}  \\
\hline
\multicolumn{5}{|c|}{\textbf{Architecture 1}}\\
\hline
\textbf{Full Precision} & 38.39 MB & 100\% & 92.35\% & \cellcolor{red!25}57.46\%\\
\textbf{BinaryNet} & 1.62 MB & 88.68\% & 86.90\% & \cellcolor{green!25}55.45\%\\
\textbf{XNOR-Net} & 1.62 MB & 87.19\% & 85.68\% & \cellcolor{green!25}51.05\%\\ 
\hline
\multicolumn{5}{|c|}{\textbf{Architecture 2}}\\
\hline
\textbf{Full Precision} & 29.83 MB & 99.34\% & 89.88\% & \cellcolor{red!25}54.86\% \\
\textbf{BinaryNet} & 0.93 MB & 97.61\% & 89.60\% & \cellcolor{green!25}54.30\%\\
\textbf{XNOR-Net} & 0.93 MB & 92.67\% & 86.68\% & \cellcolor{green!25}51.74\%\\ 
\hline
\end{tabular}
\end{center}
\label{fmnist_quantize}
\vspace{-4mm}
\end{table}

In both architectures, the computation on binarized values reduces the membership privacy risk.
However, on replacing the MAC operations with XNOR operations the membership privacy risk decreases close to random guess, but at the cost of prediction accuracy.
The corresponding results for CIFAR10 are shown in Table~\ref{cifar10quant}.

\vspace{-1mm}
\begin{tcolorbox}[text
width=0.97\columnwidth,top=1pt,bottom=1pt,left=0pt,right=0pt,
 colback=gray!10!white,colframe=gray!10!white,colbacktitle=gray!80!black]
\underline{\textbf{Summary:}} Lowering the model capacity using aggressive quantization (binarization of parameters and activation) along with XNOR computation provides strong resistance against inference attacks at the cost of accuracy.
\end{tcolorbox}
\vspace{-2mm}

\subsection{Summary of Comparison}\label{eval-summary}

We summarize the properties satisfied by each of the approaches in terms of privacy, computation, memory and energy efficiency (Table~\ref{tbl:comparison}) based on the observations in Section~\ref{eval-efficiency} and~\ref{eval-leakage}.
Note that the privacy analysis and corresponding conclusions of each of the three embedded deep learning techniques is independent of the dataset allowing us to understand their trade-offs and compare them based on the trends observed.
Here, we mark the attributes which are satisfied with $\cmark$, require additional hardware optimization as $\smark$ and do not satisfy the property with $\xmark$.

\begin{table}[!htb]
\caption{Only quantization satisfies all the efficiency and privacy requirements but suffers from accuracy degradation.}
\vspace{-3mm}
\begin{center}
\renewcommand\arraystretch{1.5}
\fontsize{7pt}{7pt}\selectfont
\begin{tabular}{|l||l|l|l|}
\hline
\textbf{Requirements} & \textbf{Compression} & \textbf{Off-the-shelf} & \textbf{Quantization} \\
\hline
\textbf{Computation Efficiency} & $\smark$    & $\xmark$ & $\cmark$  \\
\hline
\textbf{Memory Efficiency} &  $\smark$ & $\cmark$   & $\cmark$ \\
\hline
\textbf{Energy Efficiency} &  $\smark$  & $\xmark$  & $\cmark$   \\
\hline
\textbf{Privacy} &  $\xmark$   & $\cmark$   & $\cmark$  \\
\hline
\textbf{Accuracy} &  $\cmark$   & $\cmark$   & $\smark$  \\
\hline
\end{tabular}
\end{center}
\label{tbl:comparison}
\end{table}

In order to design NNs for embedded devices, quantization (binarization with XNOR computation) is an attractive design choice which not only satisfies the computation, memory and energy efficiency but also provides high resistance against membership inference attacks.
On the other hand, model compression leaks more training data membership details making it more vulnerable to membership inference attacks. Additionally, it requires hardware support and optimization to achieve better efficiency.
Off-the-shelf architectures, while provide decent privacy, do not satisfy all aspects of efficiency.
Hence, we choose quantization as a NN design choice for \method\hspace{0.01in} to provide a good three dimensional trade-off between privacy-accuracy-efficiency. We observe an accuracy degradation on using quantization and propose an approach as part of \method~ to address it.

%% file: 5design.tex
\section{\method: Design Overview}\label{design}

Based on the comparative analysis described in the previous section, we propose \method~ which answers the research question~\ref{req2}.
\method~ is a training methodology to construct NNs dedicated to embedded systems (e.g., IoT, wearable)  with efficiency, accuracy, and privacy as main requirements:

\begin{itemize}[leftmargin=*]

\item {\em Privacy.}
The model should preserve the membership privacy of an individual's data record in the training set of the model.

\item {\em Efficiency.}
The model should consume low energy, memory, and computation capacity for deployment to embedded devices.

\item {\em Accuracy.}
The model should have high test accuracy.
\end{itemize}

We address the above three requirements and reconcile the trade-offs in \method~ in two phases: In Phase I (Section~\ref{p1}), the quantized NN is generated by binarizing the parameters and activations to ensure efficiency and privacy and replacing MAC operations with XNOR computations. However, we obtain efficiency and privacy at the cost of accuracy. In Phase II (Section~\ref{p2}), we account for the accuracy and improve it by optimizing the model obtained from Phase I. We use knowledge distillation by using a state-of-the-art full precision model's predictions as target labels for training the quantized model.

\subsection{Phase I}\label{p1}

Based on our privacy-centric design space exploration in Section~\ref{eval-leakage}, we first quantize the NN's parameters and intermediate activations.
Specifically, we binarize the (parameter) values and intermediate activations (outputs) from each layer, i.e., map them to \{+1,-1\}.
This operation (as seen in Section~\ref{quant}) results in high resistance to membership inference attacks as well as satisfies the different efficiency requirements (Section~\ref{eval-efficiency}).
The NN achieves computation efficiency by replacing the expensive matrix-vector multiplications with simple Boolean arithmetic operations, i.e., XNOR computations. 
Alternatively, instead of using multiplication and addition circuits in the hardware, we leverage XNOR logic on the inputs followed by a bitcount operation (counting the number of high bits ``1'' in a binary output sequence).
The equation can be represented as follows:
$\mathbf{x} \cdot \theta =
N - 2\times\operatorname{bitcount}(\operatorname{xnor}(\mathbf{x}, \theta))$

In terms of memory efficiency, binarization results in a direct reduction of the model size as well as intermediate output memory requirements by 32x to 64x.
Lowering the precision also reduces the number of memory access by 32x to 64x resulting in a significant decrease in the energy consumption. This is due to the packing of 32x or 64x more number of single bit values for a single iteration of reading or writing in the memory compared to one value of 32 bit or 64 bit floating point values.

\begin{algorithm}
\footnotesize
\begin{algorithmic}
    \FOR{$k=1$ to $L$}
        \STATE $W_k^b \leftarrow {\rm Binarize}(W_k)$
        \STATE $a_k \leftarrow N - 2\times\operatorname{bitcount}(\operatorname{xnor}(\mathbf{a_{k-1}^b}, \mathbf{W_k^b}))$
        \IF{$k < L$}
            \STATE $a_k^b \leftarrow {\rm Binarize}(a_k)$
        \ENDIF
    \ENDFOR
\end{algorithmic}
\caption{Inference stage of binary Neural Network with XNOR operations where $W_k^b$ are the binarized weights ($W_k$) and $a_k$ is the activation of the $k^{th}$ layer}
\label{alg:inference}
\end{algorithm}

The complete inference stage of the binarized NN with XNOR computation is given in Algorithm~\ref{alg:inference}.
The matrix multiplication is done between the previous layer activation $a_{k-1}$ and the current layer's weights with the bitcount of XNOR operation's output.
Binarize() uses a threshold to map the input values to \{-1,+1\}.
In addition to the above design, we use optimizations for XNOR-Net to avoid a significant loss in accuracy.
It is well documented that it is difficult to converge a binarized model during training in case of incompatible hyperparameter settings and to this extent, we use the first and last layer of the model as full precision~\cite{AAAI1714619}.
These additional optimizations have been used previously for XNOR based networks and provide higher accuracy and model convergence at a smaller memory and energy consumption overhead~\cite{8114708,rastegari2016xnornet}.

\subsection{Phase II}
\label{p2}

While we optimize for both privacy and efficiency in Phase I (at the cost of significantly reduced accuracy shown in Section~\ref{quant}), we restore the accuracy in Phase II such that it is close to the original full precision accuracy by using knowledge distillation~\cite{44873}.
Here, we consider a pre-trained teacher model $f_{teacher}$ with state-of-the-art accuracy on the classification task and use it to guide the training of the quantized classifier from Phase-I.
During training of the quantized model (student), we do not compute the loss between the true label $y$ and prediction vector $f_{student}(x)$.
We instead estimate the cross entropy loss between the predicted label $f_{student}(x)$ and the prediction vector for the full precision teacher model $f_{teacher}(x)$.

This ensures that the student model learns to map the decision boundary of the teacher model and mimics the prediction behaviour for different inputs.
Therefore, the accuracy of the student model increases compared to the original baseline of standalone training without the teacher model. We use the same training data for the quantized model and the full precision model.

%% file: 6evaluation.tex
\section{Evaluation of \method}\label{sec:eval}

In this section, we present an extensive evaluation of \method~ on CIFAR10 dataset for Phase I (Section~\ref{evalPh1}) and Phase II (Section~\ref{evalPh2}). We focus our evaluation of \method~ on CIFAR10 to enable comparison with prior defences against membership inference attacks which are mainly evaluated on CIFAR10, e.g., differential privacy, which currently does not scale to larger datasets. Note that the proposed approach of using quantization followed by knowledge distillation is independent of the dataset and can be extended to larger datasets~\cite{44873}.

\subsection{Evaluating Phase I}\label{evalPh1}

On quantization and replacing the MACs with cheap XNOR operations, we observe that the membership inference attack accuracy decreases significantly for all the three architecture close to random guess ($\sim$50\%) (Table~\ref{cifar10quant} for CIFAR10 and Table~\ref{fmnist_quantize} for FashionMNIST).
Specifically for CIFAR10, the membership inference accuracy decreases from 56.69\% to 51.76\% for NiN, 60.40\% to 51.40\% for AlexNet and 58.70\% to 52.65\% for VGGNet.
However, since Phase I only optimizes the network for privacy and efficiency, the resultant model shows poor accuracy.
We observe a significant loss in test accuracy for all the three models: around 8\% accuracy drop from 86.16\% to 78.74\% for NiN; 14\% accuracy drop from 80.34\% to 66.8\% for AlexNet; 14\% for VGG model from 88.95\% to 74.64\%.
In order to restore the accuracy, we use knowledge distillation as described in Phase II of the \method~ training methodology.

\begin{table}[!htb]
\caption{Reducing precision lowers membership privacy risk at the cost of accuracy.}
\vspace{-3mm}
\begin{center}
\renewcommand\arraystretch{1.5}
\fontsize{7.5pt}{7.5pt}\selectfont
\begin{tabular}{|c|c|c|c|c|}
\hline
\multicolumn{5}{|c|}{\textbf{CIFAR10}} \\
\hline
\multicolumn{2}{|c|}{\textbf{Architecture}} & \textbf{Train}  & \textbf{Test}  & \textbf{Inference}  \\
 \multicolumn{2}{|c|}{} & \textbf{Accuracy} & \textbf{Accuracy} & \textbf{Accuracy}  \\
\hline
\multirow{2}{*}{\textbf{NiN}} & \textbf{Full Precision} & 98.16\% & 86.16\% & \cellcolor{red!25}56.69\% \\
& \textbf{Binary Precision} & 81.93\% & 78.74\% & \cellcolor{green!25}51.76\% \\
\hline
\multirow{2}{*}{\textbf{AlexNet}} & \textbf{Full Precision} & 97.86\% & 80.34\% & \cellcolor{red!25}60.40\% \\
& \textbf{Binary Precision} & 68.62\% & 66.8\% & \cellcolor{green!25}51.40\% \\
\hline
\multirow{2}{*}{\textbf{VGG13}} & \textbf{Full Precision} & 99.58\% & 88.95\% & \cellcolor{red!25}58.70\%\\
& \textbf{Binary Precision} & 79.67\% & 74.64\% & \cellcolor{green!25}52.65\%\\
\hline
\end{tabular}
\end{center}
\label{cifar10quant}
\vspace{-3mm}
\end{table}

The privacy provided by quantized NN is due to the decrease in overfitting, empirically measured as the difference between the train and test accuracy.
Furthermore, \method~ models have a lower model capacity on quantizing the parameters which lowers the memorization of training data by the parameters.
Further, the quantization acts as a noise to strongly regularize the model~\cite{NIPS2016_6573}.
At the same time, this optimization provides high degree of efficiency to be executed on low powered embedded devices.

\subsection{Evaluating Phase II}\label{evalPh2}

The objective of Phase II of \method~ is to enhance the accuracy of the quantized model with XNOR computations which depicts high membership inference attack resistance and efficiency.
In Phase II, we use the teacher-student model (described in Section~\ref{design}) to train the quantized student model. The quantized model training is supervised using the output predictions of the full precision teacher model.
Here, Phase II is heterogeneous, i.e., we are flexible to choose any full precision teacher model which can provide high accuracy on the considered CIFAR10 dataset (Table~\ref{kd}).
For the teacher models, we consider pre-trained state-of-the-art architectures\footnote{https://github.com/huyvnphan/PyTorch\_CIFAR10}: DenseNet169 and ResNet50, along with the full precision versions of NiN, Alexnet and VGGNet.
The standalone test accuracy of the DenseNet169 and ResNet50 architectures are 92.84\% and 92.12\% respectively with membership inference attack accuracy of ~55\%. The full precision test and attack accuracy for NiN, AlexNet and VGGNet are given in Table~\ref{cifar10quant}.

\setlength\tabcolsep{1pt}
\begin{table}[!htb]
\caption{Phase II of \method~ improves the accuracy of the private-efficient model from Phase I (CIFAR10).}
\vspace{-3mm}
\begin{center}
\renewcommand\arraystretch{1.5}
\fontsize{7.5pt}{7.5pt}\selectfont
\begin{tabular}{|c|c|c|c|c|c|}
\hline
\textbf{Teacher} & \textbf{Student} & \textbf{Train}  & \textbf{Test}  & \textbf{Inference}  \\
&  & \textbf{Accuracy} & \textbf{Accuracy} & \textbf{Accuracy}  \\
\hline
\multicolumn{5}{|c|}{\textbf{Standalone Models}}\\
\hline
\textbf{Binary NiN} & - & 81.93\% & 78.74\% & 51.76\% \\
\textbf{Binary AlexNet} & - & 68.62\% & 66.8\% & 51.40\% \\
\textbf{Binary VGG13} & - & 79.67\% & 74.64\% & 52.65\%\\
\hline
\multicolumn{5}{|c|}{\textbf{Homogeneous Architecture Distillation}}\\
\hline
\textbf{NiN} & \textbf{Binary NiN} & 90.49\% & 83.52\% & 53.90\% \\
\textbf{AlexNet} & \textbf{Binary AlexNet} & 76.79\% & 73.5\% & 51.85\% \\
\textbf{VGG13} & \textbf{Binary VGG13} & 89.45\% & 81.58\% & 54.98\%\\
\hline
\multicolumn{5}{|c|}{\textbf{Heterogeneous Architecture Distillation}}\\
\hline
\textbf{DenseNet169} & \textbf{NiN} & 92.84\% & 83.71\% & 54.95\%\\
\textbf{DenseNet169} & \textbf{AlexNet} & 81.87\% & 76.23\% & 53.51\%\\
\textbf{DenseNet169} & \textbf{VGG13} & 93.45\% & 85.8\% & 54.17\%\\
\hline
\textbf{ResNet50} & \textbf{NiN} & 91.74\% & 83.77\% & 54.53\% \\
\textbf{ResNet50} & \textbf{AlexNet} & 80.12\% & 74.92\% & 53.12\%\\
\textbf{ResNet50} & \textbf{VGG13} & 94.23\% & 86.52\% & 54.46\%\\
\hline
\end{tabular}
\end{center}
\label{kd}
\vspace{-4mm}
\end{table}

\noindent\textbf{Homogeneous distillation} transfers knowledge from the same full precision model architectures to their quantized versions, e.g., full precision NiN with Binarized NiN.
Here, we see that there is 5\% increase in test accuracy (from 78.74\% reported Table~\ref{cifar10quant} to 83.52\%) for NiN with an increase of 2\% in membership inference attack.
Similarly, there is an increase of 7\% test accuracy for AlexNet with a very small membership privacy leakage increase of 0.45\%; and increase of 7\% test accuracy at the cost of 2\% membership inference attack accuracy for VGGNet.

\noindent\textbf{Heterogeneous distillation} combines different architectures, e.g., DenseNet169 and ResNet50, with the quantized models from Phase I. We see that the increase in test accuracy is only minimally higher than the homogeneous models for NiN and AlexNet but shows a higher increase in the membership inference attack accuracy.
However, in case of VGGNet, we observe an increase of 4\% additional test accuracy compared to homogeneous knowledge distillation with a small decrease in the membership inference test accuracy.
In Phase II, increase in test accuracy is accompanied with a small but acceptable increase in the inference attack accuracy indicating a privacy-utility trade-off.
Compared to the full precision counterparts, we observe that the distilled models show an accuracy degradation of only 3\% for NiN (86.66\% to 83.77\%), 4\% for AlexNet (80.34\% to 76.23\%) and 2\% for VGGNet (88.95\% to 86.52\%).

\vspace{-1mm}
\begin{tcolorbox}[text width=0.97\columnwidth,top=1pt,bottom=1pt,left=0pt,right=0pt,
 colback=gray!10!white,colframe=gray!10!white,colbacktitle=gray!80!black]
\underline{\textbf{Summary:}} Compared to heterogeneous distillation, homogeneous distillation results in higher improvement in test accuracy but small gain in privacy. In view of this privacy-accuracy trade-off, the choice of using homogeneous or heterogeneous knowledge distillation is specific to the architecture and the acceptable privacy-accuracy requirements of the application.
\end{tcolorbox}

\begin{figure}[!htb]
    \centering
    \subfloat[Without Gecko]{{\includegraphics[width=0.23\textwidth]{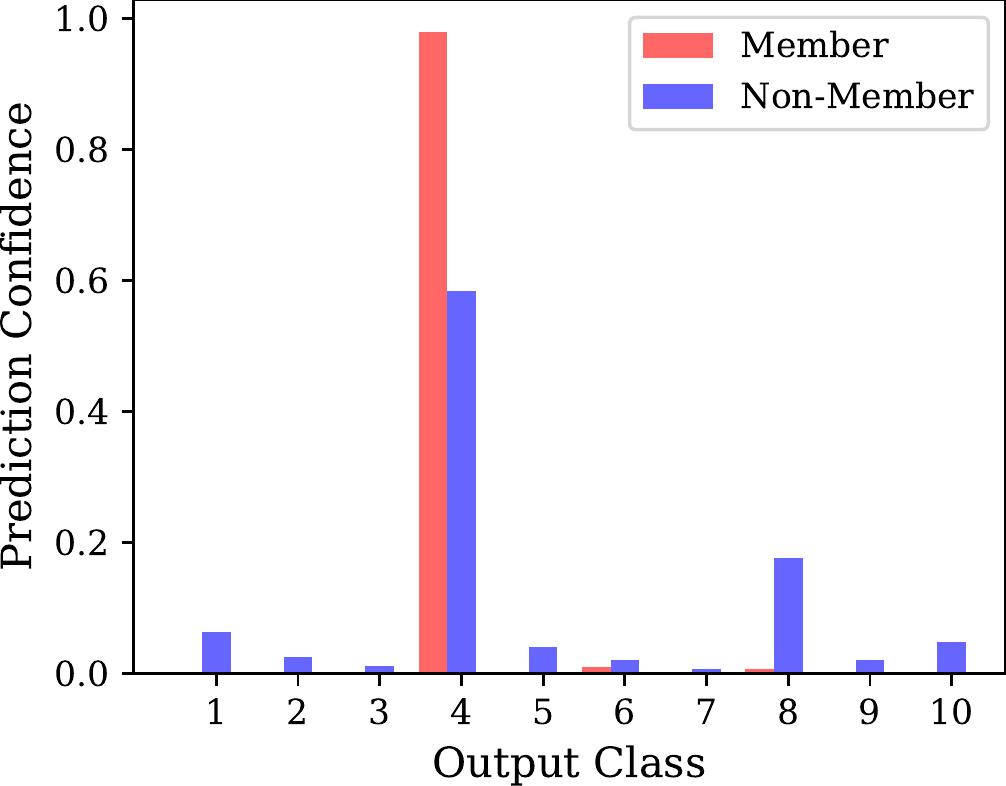} }}%
    \subfloat[With Gecko]{{\includegraphics[width=0.23\textwidth]{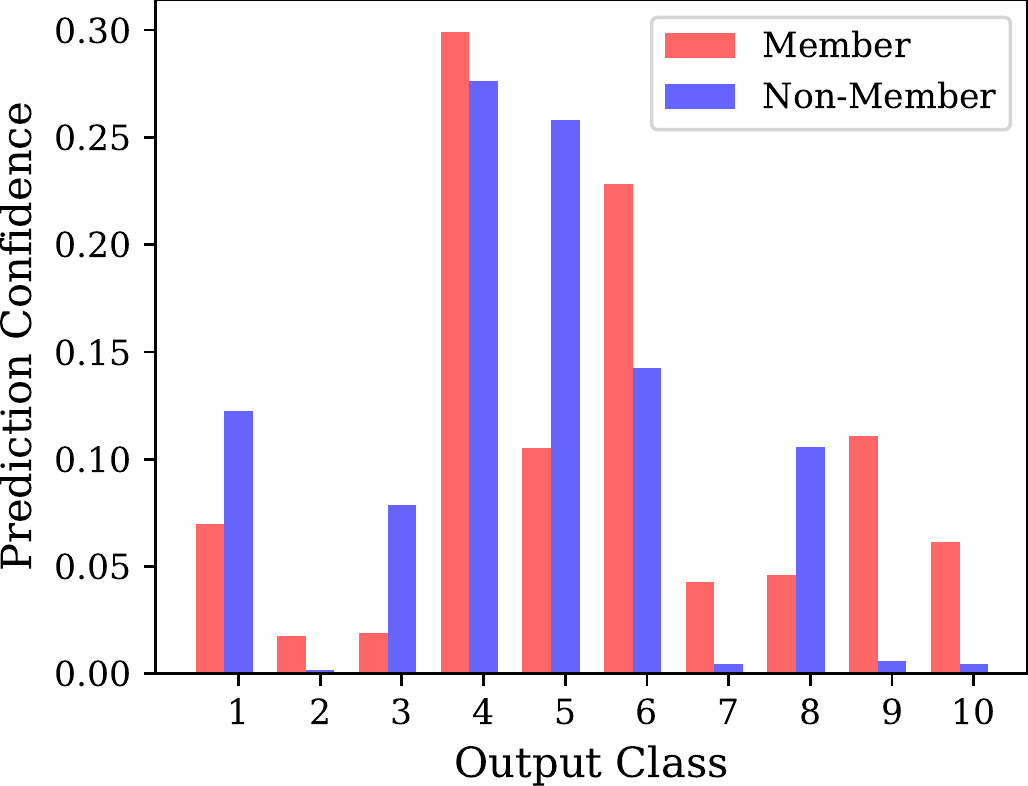} }}%
\caption{(a) Distinguishable predictions between train and test data records in undefended models makes them vulnerable to membership inference attacks, (b) \method~ models have indistinguishable confidence scores (CIFAR10).}
\label{fig:gecko}
\end{figure}

\noindent\textbf{Explaining the privacy gain.} The \method~ framework results in models which make the output confidence of the train and test data records similar reducing the inference attack accuracy (Figure~\ref{fig:gecko}).

\begin{figure}[!htb]
\includegraphics[width=6cm]{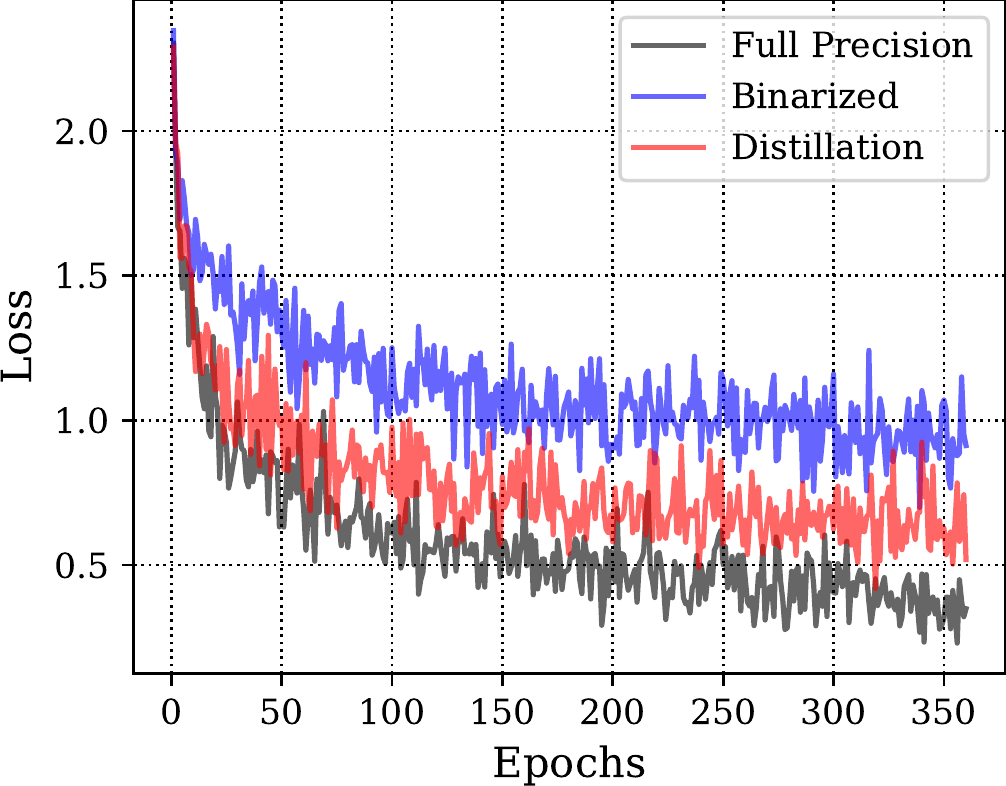}
\caption{Loss functions in Phase I (Binarized) are higher than Phase II (Distilled Binarized) indicating the improvement in accuracy (CIFAR10).}
\label{fig:loss}
\end{figure}

Further, knowledge distillation in Phase II enables to lower the model's loss compared to Phase I resulting in higher test accuracy as shown in Figure~\ref{fig:loss}.
However, this loss function is still higher than the full precision version indicating the small test accuracy degradation for some privacy gain.

\section{Comparison with Prior Defences}\label{eval-defences}

The privacy defences proposed in literature can be categorized into (a) modification of training algorithm (e.g., adversarial regularization (AR) and differential privacy (DP)) and (b) post-training techniques (e.g., MemGuard).
\method~ training methodology is part of category (a) where we modify the training of the ML model in order to provide acceptable levels of privacy and accuracy.

\noindent\textbf{\underline{MemGuard~\cite{10.1145/3319535.3363201}}} adds carefully crafted noise to the target model's output observations to ensure the misclassification of the adversary's attack model). MemGuard has been shown to be ineffective against the membership inference attack considered in this work, i.e., prediction confidence attack~\cite{song2020systematic}. Hence, we consider AR and DP as baselines for comparison.

\noindent\textbf{\underline{Adversarial Regularization (AR)~\cite{DBLP:conf/ccs/NasrSH18}.}} The problem of defending against membership inference attack is modelled as a minimax game between two NNs: classifier model and attacker model.
The two models are trained alternatively with conflicting objectives: first, the attacker model is trained to distinguish between the training data members and non-members followed by training the classifier model to minimize the loss as well as fool the attacker model.

\noindent\textbf{\underline{Differential privacy (DP)~\cite{Abadi:2016:DLD:2976749.2978318}.}} We consider DP-SGD which adds carefully crafted noise to the gradients during backpropagation in stochastic gradient descent algorithm.
The noise is sampled from a Laplacian or Gaussian distribution proportional to the model's sensitivity which is then added to the gradients during backpropagation.
This provides provable bound on the information leaked about an individual data record in the dataset and ensures that the presence or absence of a data record does not change the model's output, hence defending against membership inference attacks.

\begin{figure}[!htb]
\includegraphics[width=8cm]{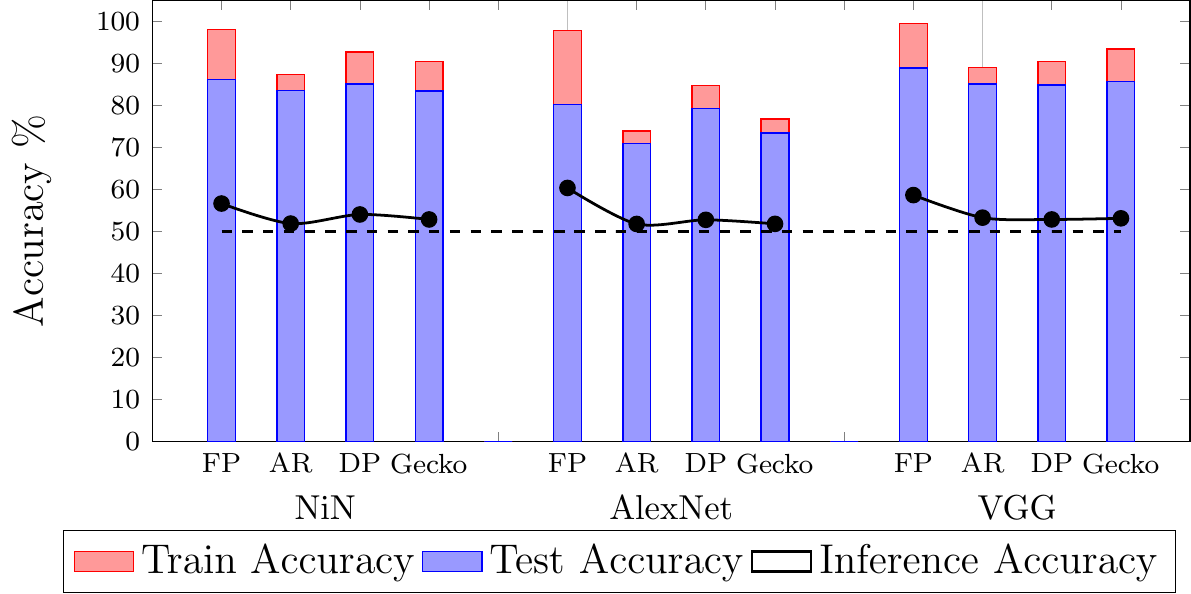}
\caption{\method~ models are comparable to prior state-of-the-art privacy defences in terms of test accuracy and inference accuracy while additionally ensuring efficiency (CIFAR10).}
\label{fig:compare}
\end{figure}

\noindent\textbf{Results.} The comparison of models trained using \method~ is shown in Figure~\ref{fig:compare} (FP stands for Full Privacy and reports results without defence).
Models trained using \method~ are comparable in test accuracy and resisting membership inference leakage to AR and DP.
The inference accuracy for NiN is 52.90\% (\method) compared to 54.09\% (DP) and 51.92\% (AR) and test accuracy of 83.52\% (\method) compared to 85.11\% (DP) and 83.66\% (AR).
For AlexNet, the inference accuracy is 51.85\% (\method) compared to 52.81\% (DP) and 51.83\% (AR) and test accuracy of 73.5\% (\method) compared to 79.27\% (DP) and 71.02\% (AR).
For VGGNet, the inference accuracy is 53.17\% (\method) compared to 52.90\% (DP) and 53.33\% (AR) and test accuracy of 85.8\% (\method) compared to 84.91\% (DP) and 85.19\% (AR).\\

\begin{tcolorbox}[text width=0.97\columnwidth,top=1pt,bottom=1pt,left=0pt,right=0pt,
 colback=gray!10!white,colframe=gray!10!white,colbacktitle=gray!80!black]
\underline{\textbf{Summary:}} \method~ has comparable accuracy and privacy to prior defences against prediction confidence based membership inference attack. Unlike other defences, \method~ is efficient for deployment to embedded systems.
\end{tcolorbox}

%% file: 7related.tex
\section{Related Work}\label{related}

\noindent\textbf{\underline{Embedded Deep Learning.}} Several works have explored optimization of NNs via quantization to different precision~\cite{Hubara:2017:QNN:3122009.3242044,NIPS2016_6573,rastegari2016xnornet,Li2016TernaryWN} and the challenges in training them~\cite{AAAI1714619}. Further, specially designed NNs with low memory footprint (e.g., SqueezeNet and MobileNet) have been deployed for low powered devices such as mobile phones and micro-controllers~\cite{DBLP:journals/corr/IandolaMAHDK16,conf/cvpr/SandlerHZZC18}. 
In addition to traditional model compression techniques considered in this work~\cite{DBLP:journals/corr/HanPNMTECTD16,Han:2015:LBW:2969239.2969366,DBLP:journals/corr/HanMD15}, recent work prune the model with specific objective functions (e.g., efficiency)~\cite{journals/corr/YangCS16a}. Alternatively, hardware accelerators are designed to reuse weights and intermediate computation which enable significant performance improvement~\cite{10.1109/ISCA.2016.30} and optimized specifically for low precision NNs~\cite{Umuroglu2017FINNAF} while skipping computations to improve performance for compressed models~\cite{9043731}. 
However, none of these work consider privacy as a design choice and address the efficiency, accuracy, and privacy trade-off.

\noindent\textbf{\underline{Privacy in Deep Learning.}} Privacy risks in ML models have been extensively studied via inference attacks such as membership inference in a blackbox setting~\cite{salem2018ml,shokri2017membership} or in the context of whitebox setting~\cite{DBLP:journals/corr/abs-1812-00910}.
Further, membership privacy risks have been shown for generative models~\cite{LOGANMembershipInferenceAttacksAgainstGenerativeModels}, graph models~\cite{duddu2020quantifying} and federated learning~\cite{melis2019exploiting,DBLP:journals/corr/abs-1812-00910}.
Other than membership inference attacks, attribute/property inference attacks enable an adversary to infer sensitive attributes from the training dataset~\cite{melis2019exploiting,10.1145/3243734.3243834}. However, none of the prior work consider the privacy risks in the context of embedded deep learning.

%% file: 8conclusions.tex
\section{Conclusions}\label{conclusions}

We perform a privacy-centred design space exploration of state-of-the-art algorithms for improving efficiency in NNs: model compression, quantization and efficient off-the-shelf architectures. We identify quantization as a design choice which shows high resistance against membership inference attacks while satisfying all the efficiency requirements. Model compression, after retraining to restore accuracy, leaks more membership information compared to the original uncompressed model while off-the-shelf architectures do not provide the best efficiency guarantees.
Based on our observation, we propose a two phase \method~ training framework to design private, efficient and accurate NNs for deployment to low powered embedded devices.